\documentclass[conference, letter]{IEEEtran}
\IEEEoverridecommandlockouts
\usepackage{graphicx} 
\usepackage{amsmath}  
\usepackage{amssymb}
\usepackage{color}
\usepackage{multirow}
\usepackage[space]{cite}

\usepackage{amsmath}
\usepackage{amssymb}

\title{Floating-Point Multiply-Add with Approximate Normalization for Low-cost Matrix Engines}
\author{
\IEEEauthorblockN{Kosmas Alexandridis, Christodoulos Peltekis, Dionysios Filippas and Giorgos Dimitrakopoulos}
\IEEEauthorblockA{Electrical and Computer Engineering, Democritus University of Thrace, Xanthi, Greece
\thanks{This work was supported by a research grant of Siemens EDA to Democritus University of Thrace for ``HLS Research for Systems-on-Chip''.}}}

\begin{document}

\maketitle

\begin{abstract}
The widespread adoption of machine learning algorithms necessitates hardware acceleration to ensure efficient performance. This acceleration relies on custom matrix engines that operate on full or reduced-precision floating-point arithmetic. However, conventional floating-point implementations can be power hungry. This paper proposes a method to improve the energy efficiency of the matrix engines used in machine learning algorithm acceleration. Our approach leverages approximate normalization within the floating-point multiply-add units as a means to reduce their hardware complexity, without sacrificing overall machine-learning model accuracy. Hardware synthesis results show that this technique reduces area and power consumption roughly by 16\% and 13\% on average for Bfloat16 format. Also, the error introduced in transformer model accuracy is 1\% on average, for the most efficient configuration of the proposed approach.
\end{abstract}

\section{Introduction}
\begin{figure}[t]
\centering
\includegraphics[width=0.9\columnwidth]{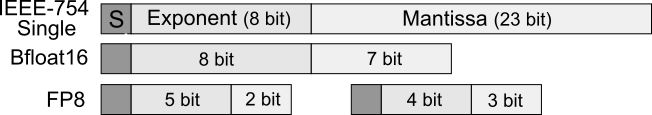}
\caption{Floating point formats: Single and reduced precision.}
\label{f:flp-formats}
\end{figure}

The ever-expanding presence of machine learning algorithms across diverse application domains necessitates hardware acceleration to address performance and energy efficiency bottlenecks~\cite{jouppi}. While floating-point arithmetic remains dominant for data representation due to its accuracy, recently proposed reduced-precision formats like Bfloat16 and FP8 offer an attractive middle ground. They balance the precision of floating-point with the hardware efficiency of integer arithmetic~\cite{bfloat-def, tensorfloat}. Fig.~\ref{f:flp-formats} illustrates the format of various widely-used floating-point formats.

Hardware acceleration for machine learning applications often relies on custom matrix engines that leverage full or reduced-precision floating-point arithmetic. These engines utilize systolic arrays~\cite{kung,intelMPU}, characterized by their regular structure and data flow. The building blocks of these arrays are processing elements that perform the essential multiply-add operations for matrix multiplication.

This work aims at improving the energy efficiency and area footprint of such matrix engines by introducing appropriate
floating-point based matrix engines by introducing 
approximations during arithmetic calculations that simplify hardware  
without compromising the overall accuracy of the underlying machine learning models.

More specifically, prior research in the early-days of floating-point arithmetic, observed a strong dependency between the number of normalization shifts, the exponent differences and signs of the input operands in floating point addition~\cite{fields, oberman}. In fact, it was shown that massive normalization shifts are very rare and can occur only under very specific circumstances. In this work, we leverage this observation and design floating-point multiply-add units that employ \emph{approximate normalization}. The result of a floating point adder is always normalized by three possible fixed shift amounts that does not depend on the count of leading zeros/ones of the result.
Even if fixed-shift-amount normalization may leave in rare cases the result un-normalized, it does not alter significantly the accuracy of machine-learning applications.

This argument was tested by executing representative machine-learning applications for the BERT transformer~\cite{bert} 
using matrix engines that employ accurate and approximate normalization in their floating-point datapaths.
Experimental results show that employing approximate normalization reduces area and power consumption of the floating-point datapaths by 16\% and 13\% on average without introducing substantial errors that would degrade the final accuracy of the floating-point machine learning models. In particular, approximate normalization degraded relevant accuracy metrics of the examined transformer models by 1\%--7\% on average, depending on the examined configuration.

\section{Organization of floating-point matrix engines}
Machine learning algorithms, for both training and inference tasks, rely mainly on matrix multiplication operations. Matrix engines are often built using systolic arrays that offer a compelling solution for hardware acceleration of matrix multiplication. These arrays consist of a grid of identical processing elements (PEs) that communicate with their neighbours. This regular structure simplifies the design process of such architectures and facilitates efficient data flow patterns used for different application cases. 

\begin{figure}[t]
\centering
\includegraphics[width=0.9\columnwidth]{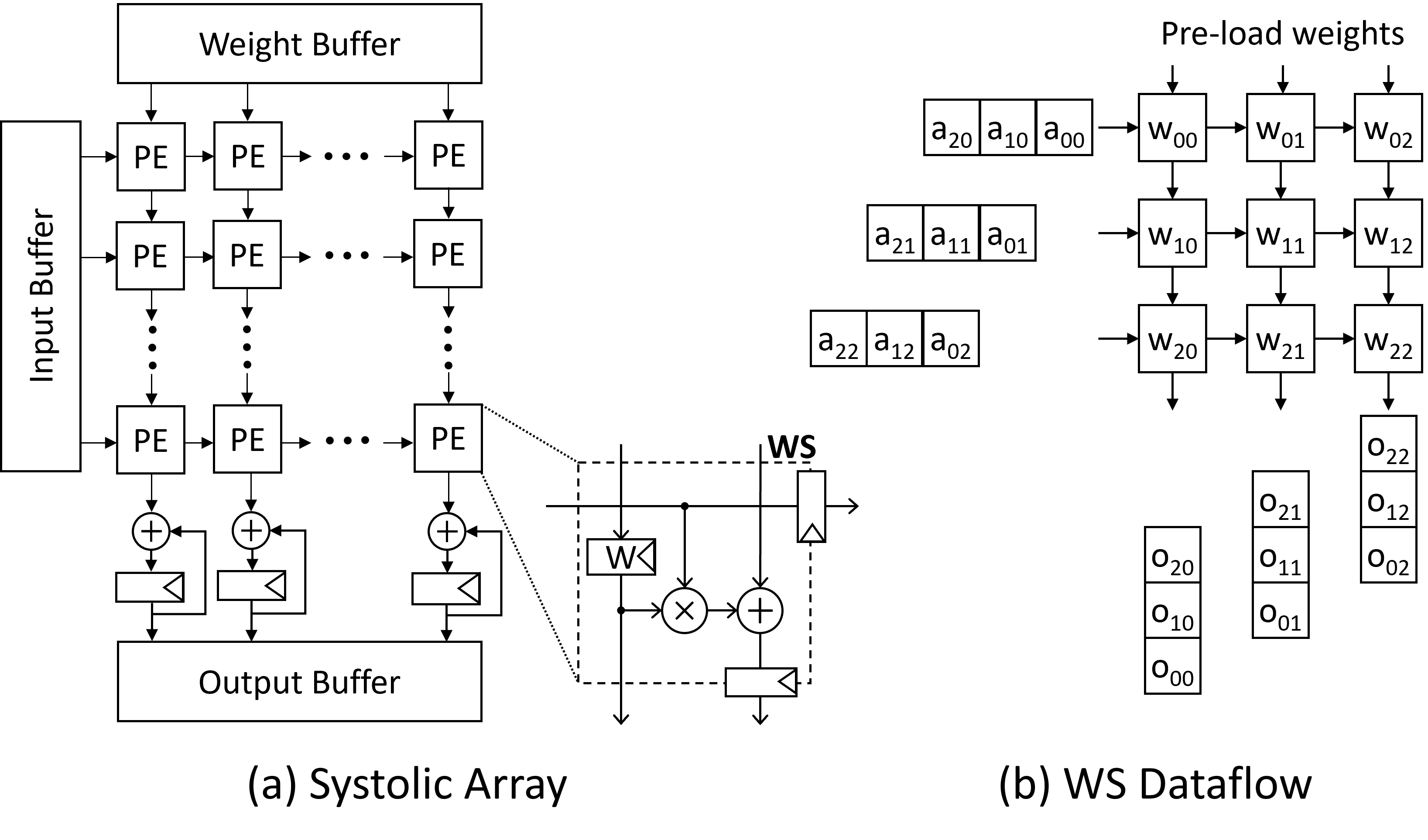}
\caption{a) The organization of systolic arrays and b) the corresponding weight-stationary dataflow.}
\label{f:sys_arch}
\end{figure}

Systolic arrays (SAs) can be programmed to support various dataflows, with weight-stationary being a popular choice. In this approach, weights are first loaded into the PEs from the north side. Then, the input data is streamed, from west to east, across the array. As the data flows, each PE performs a multiplication between its locally stored weight and the incoming input. The product is then added to a partial sum received from the north input of the PE. This accumulation process continues downwards in the vertical direction, with the final results emerging at the south end of the systolic array as shown in Fig. \ref{f:sys_arch}. 

Leveraging fused multiply-add units within each PE further optimizes the matrix engine design. These units combine multiplication and addition into a single operation, eliminating the need for intermediate rounding steps within the PEs~\cite{lateff_fma, exp_an_mmfus}. Normalization still occurs after each addition to prevent a large number of significand (mantissa plus one hidden bit)  bits from being lost in subsequent alignment stages of PEs, which would ultimately lead to an inaccurate final result. As shown in~\cite{fpadd_mon}, normalization at each step is also needed to preserve the monotonicity of multi-term addition. 
On the contrary, to reduce hardware overhead, rounding can occur only once at the south end of each column. This approach necessitates keeping a higher bit width for all intermediate addition results within the PEs to prevent precision loss~\cite{skewed_systolic}. 

\begin{figure}[t]
\centering
\includegraphics[width=0.7\columnwidth]{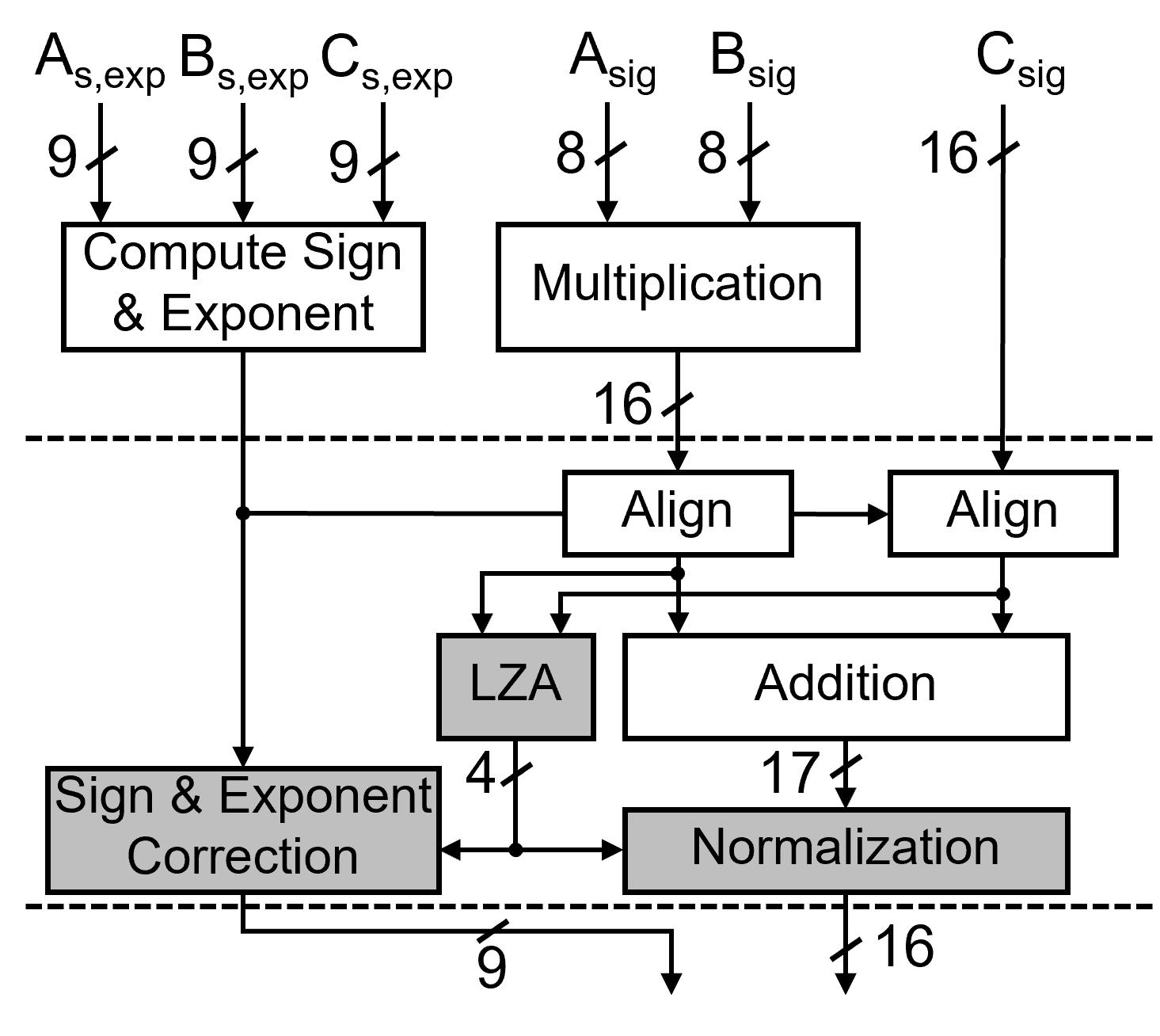}
\caption{The pipelined structure of a fused multiply-add PE for reduced-precision Bfloat16 numbers that uses accurate normalization logic that includes Leading-Zero Anticipation and counting (LZA), the normalization shifter and the sign and exponent correction logic. Significands of $A$ and $B$ (7 mantissa bits plus one hidden bit) are 8-bit. The significand of partial sum $C$ and the output assume double bitwidth of 16 bits.}
\label{f:mul-add-pe}
\end{figure}

As illustrated in Fig.~\ref{f:mul-add-pe} each PE is split into two pipeline stages. In the first stage, the significands of $A$ and $B$ are multiplied. In parallel, their exponents are added and the sum of exponents is compared to the exponent of the partial sum $C$. Their exponent difference determines on how the product of $A\times B$ and $C$ should be aligned before being added in the subsequent pipeline stage. 
In the second pipeline stage, alignment and addition take place. Leading-Zero Anticipation and counting (LZA)~\cite{lza, lp_lza} that operates in parallel to addition, predicts the required shift amount to normalize the adder's output. This shift value is also used to adjust the previously computed exponent of the final result. 

This pipeline organization is preferred for reduced-precision floating-point arithmetic.
In the past, where single or even double-precision floating point formats were employed, the delay of the multiplier was so large that it could hide the delay of exponent addition and comparison as well as the alignment of partial sum $C$. In the case of reduced-precision arithmetic, this is not beneficial and alignment is scheduled in the second pipeline stage.

\begin{figure}[t]
\centering
\includegraphics[width=0.6\columnwidth]{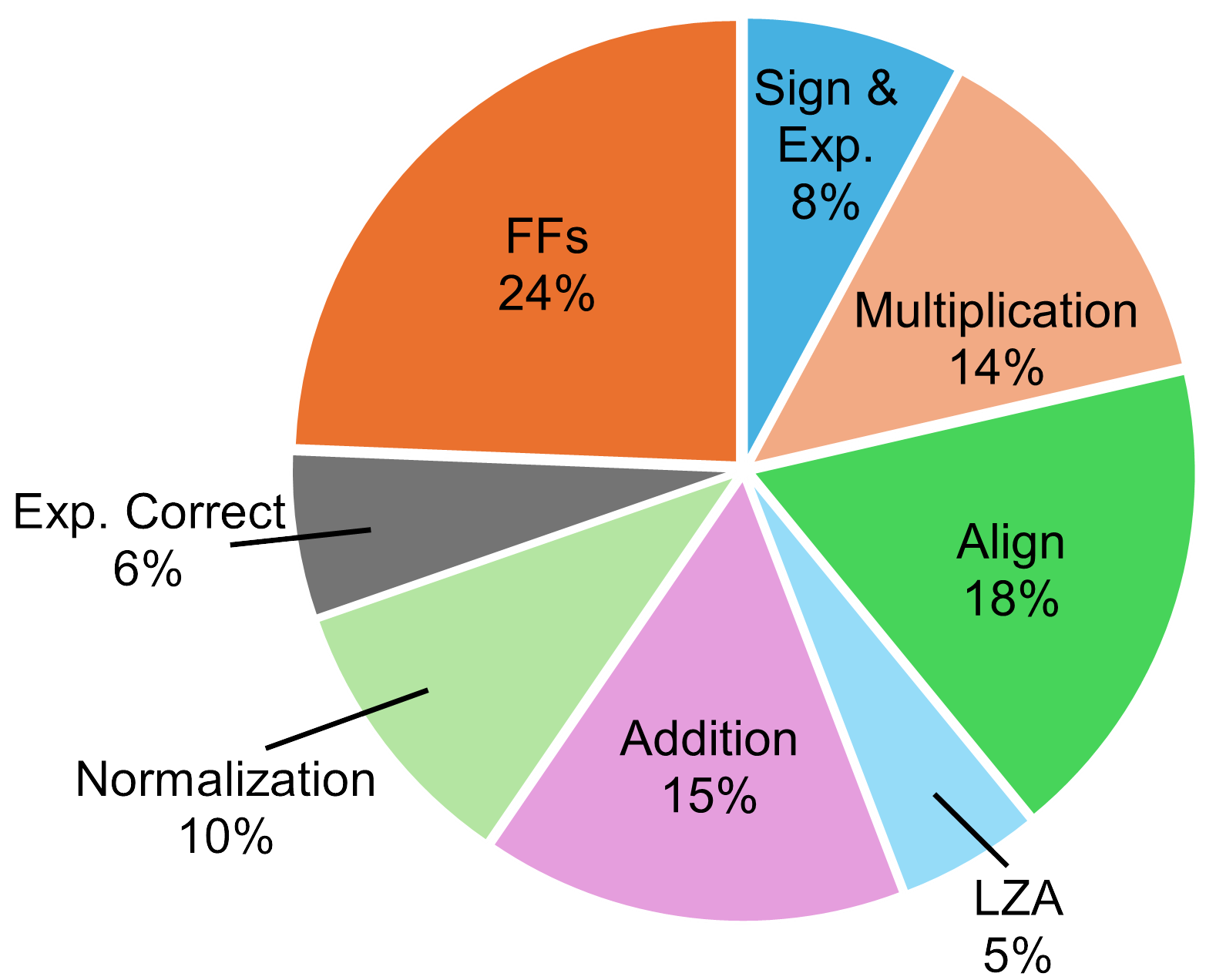}
\caption{The area contribution of each major component of the floating-pointing PE of Fig~\ref{f:mul-add-pe}. Inputs and outputs follow Bfloat16 representation. However, partial sum $C$ and the output of the PE assume 16-bit significands for preserving precision in the addition step executed in each column of the SA. Flip-flops (FFs) refer to the area contribution of the pipeline registers.}
\label{f:area-pie-bfloat16}
\end{figure}

Fig.~\ref{f:area-pie-bfloat16}  depicts the percentage of area occupied by each major component of a floating-point PE assuming a Bfloat16 representation for the input operands $A$ and $B$: 8-bit exponents and 8-bit significands (7-bit mantissa plus one hidden bit). Partial sum $C$ and the output of the PE has the same 8-bit exponent but use a 16-bit significant. This is done to accommodate higher precision for the per-column addition imposed by the structure of the SA. At the south end of the SA, a rounding module returns the final output to a typical Bfloat16 number. 
The area contribution of each component is taken after logic synthesis at 28nm targeting a clock frequency of 1GHz.

\section{Floating-Point Processing Elements with Approximate Normalization}

To reduce the hardware cost of a floating-point PE, we focus on replacing accurate normalization 
represented by the dark-gray components in Fig. \ref{f:mul-add-pe}, with an \emph{approximate normalization} step. Approximate normalization means that we don't need to count (or equivalently predict) exactly the number of leading zeros in the sum. Instead, we perform a simpler check on some of the sum's bits and based on the result we perform approximate normalization by shifting the sum by a fixed (and predetermined) number of positions.

We selected this approach for two reasons:
First, normalization logic covers a substantial percentage of the total area of each PE, e.g., the area of all relevant components shown in Fig. \ref{f:area-pie-bfloat16} sum up to 21\% of the total area of each PE.
Second, approximating normalization, has a minimal impact on the numerical accuracy of the multiply-add step done in each PE. On the contrary, if we employed approximate multiplication or addition we could have saved more area but with the risk to potentially cause undesirable amount of error. 

\subsection{How much normalization is needed per PE?}
In each PE, the product of the input and the weight stored locally in that PE is added to the partial sum produced by the PE in the same column but in the previous row.
Based on the decades-old insights provided in~\cite{fields, oberman}, the amount of normalization needed after addition depends on the signs of the two addends and on their exponent difference.

When operands of addition have the same sign the effective operation is always an addition of their absolute values and the result's sign is the common sign of the two inputs. In this case, normalization involves either a 1-bit right shift when addition overflows or no-shift at all~\cite{fields}.

In the case of unlike signs, the effective operation when adding two numbers is a subtraction. 
The sign of the result is the same as the sign of the operand with the largest absolute value. Subtraction may cause a large number of leading zeros in the result. There are three distinct cases to consider according to the exponents of the two addends: a) When the exponents are equal; b) when they differ by exactly one; and c) when their difference is larger than one.
Cases (a) and (b) are the only ones where there is chance to end up with more than one leading zero in the final sum. For case (c), only a single leading zero can occur.

Also, the probability that a large normalization shift needed in cases a) and b) is very low 
as proven formally in~\cite{fields, oberman} and observed also by our measurements in the experimental results. This property has been used for many years in the design of floating-point units of high-end processors. The corresponding floating point units followed a dual path architecture~\cite{dual-path} where addition is computed with or without normalization. The path that executed the addition was decided according to the sign of the input operands and their exponent difference. 

\subsection{Approximate Normalization}
In this work, we leverage the fact that large normalization shifts are rarely needed and design a low-cost approximate normalization architecture for the floating-point multiply-add units found in matrix engines.

Instead of detecting leading zeros in the full width of an addition result, only a fixed amount of most significant bits is examined. To compute the normalization of the sum's result we compute the OR of the first $k$ most significant bits (OR-reduce) and the OR of the following $\lambda$ most significant bits, to detect if either of them contains a $1$ and then perform the corresponding normalization shift. This leads to three possible outcomes:

\begin{itemize}
  \item If the $k$ most significant bits contain at least a single $1$, no normalization shift happens regardless of second check's result.
  \item If the $k$ most significant bits are all $0$ and the next $\lambda$ bits contain a $1$, then the sum is shifted $k$ bits to the left.
  \item If both checks failed to detect a 1 then the sum is shifted $k+\lambda$ bits to the left. 
\end{itemize}

\begin{figure}
\centering
\includegraphics[scale=0.6]{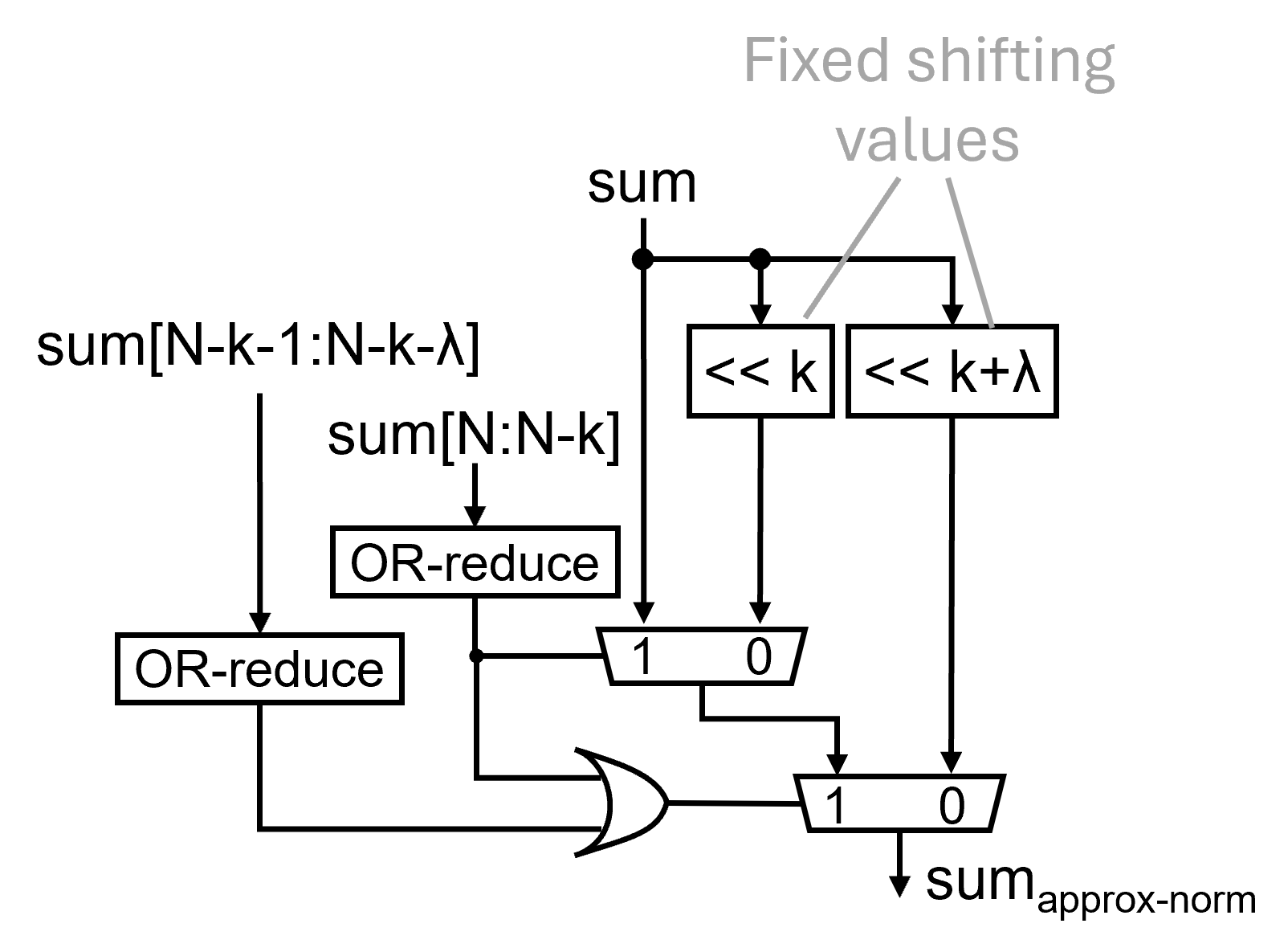}
\caption{The logic that implements approximate normalization. The result of the addition ('sum') is either not shifted, or shifted for $k$ or $k+\lambda$ positions.}
\label{f:k-bit_check_arch}
\end{figure}

As shown in Fig. \ref{f:k-bit_check_arch}, the hardware needed to perform the bit check and the subsequent shifting involves a $k$-term and a $\lambda$-term parallel OR-reduction trees for calculating the shift amount and two levels of 2-to-1 multiplexing that perform the constant shift-amount normalization, controlled by the results of the OR-trees. 
The same conditions that drive the two multiplexers, also control the update of the exponent. 

Approximate normalization may sometimes produce un-normalized results. When the result is only partially normalized subsequent alignment steps may shift out a number of bits that would otherwise contribute to the accuracy of the final result. However, the low probability of needing large normalization shifts and the double-precision of the partial sums mitigate the induced error.

\begin{table*}[t!]
    \centering
    \caption{Performance of BERT model for 10 benchmarks of the GLUE dataset~\cite{glue}. }
    \begin{tabular}{|c|c||c|c|c|c|c|c|c|c|c|c|}
        \hline
          \multicolumn{2}{|c||}{Benchmarks} &{\bf STS-2}&{\bf MNLI-m}& {\bf MNLI-mm}&{\bf QQP}&{\bf QNLI}&{\bf CoLA}&{\bf MRPC}& {\bf RTE}& {\bf WNLI}& {\bf STS-B}  \\\hline \hline
        \multirow{3}{*}{Accuracy (\%)}
        &FP32        & 92.1 & 79.2   & 84.2  &93.1 & 93.3 & 53.6 & 86   & 74.3 & 56.3 & 92 \\
        &BF16        & 93.1 & 80     & 83.3  &93.1 & 93.3 & 53.6 & 86   & 74.3 & 56.3 & 92 \\
        &BF16an-1-1  & 92.1 & 78.2   & 83.3  &93.1 & 92.2 & 52.6 & 86   & 73.4 & 54.9 & 92 \\
        &BF16an-1-2  & 92.1 & 79.2   & 83.3  &93.1 & 91   & 52.6 & 86   & 73.3 & 56.3 & 92 \\
        &BF16an-2-2  & 90   & 69.3   & 77    &86.1 & 81.2 & 57.4   & 79.2 & 64.4 & 52.1 & 87.3 \\
        \hline \hline
        \multirow{3}{*}{F1-score}
        &FP32        &0.921 & 0.794  & 0.845  &0.93 & 0.933 & 0.424 & 0.857 & 0.726 & 0.406 & - \\
        &BF16        &0.931 & 0.8    & 0.833   &0.93 & 0.920 & 0.424 & 0.859 & 0.738 & 0.406 & - \\
        &BF16an-1-1  &0.921 & 0.784  & 0.833  &0.93 & 0.922 & 0.413 & 0.859 & 0.726 & 0.399 & - \\
        &BF16an-1-2  &0.921 & 0.794  & 0.833  &0.93 & 0.910 & 0.413 & 0.859 & 0.715 & 0.406 & - \\
        &BF16an-2-2  &0.900 & 0.690   & 0.770   &0.86 & 0.810 & 0.560  & 0.79  & 0.644 & 0.407 & - \\
        \hline
    \end{tabular}
    \label{t:accuracy}
\end{table*}

\section{Evaluation}
\label{s:results}
Experimental evaluation aims to examine the impact of approximate normalization on real machine-learning applications and highlight the achieved hardware savings. 

\subsection{Accuracy of BERT using approximate normalization}
To quantify the accuracy loss induced by approximate normalization vs accurate normalization, we experimented on the BERT Transformer model~\cite{bert} for ten different benchmarks of the GLUE dataset~\cite{glue}. The results are shown in Table \ref{t:accuracy}. 
We report the Accuracy and F1-score for all benchmarks with an exception for STS-B, where we report the Pearson Correlation Coefficient (PCC).

We examined three cases: a) An FP32 model which computes every matrix multiplication in single precision;
b) a BF16 model that uses reduced-precision Bfloat16 format with accurate normalization for multiply-add operations; and c) a BF16an-$k$-$\lambda$ model which uses Bfloat16 arithmetic but employs approximate normalization for all multiply-add operations. 
For instance, BF16an-1-2 corresponds to the case that approximate normalization checks $k=1$ and $\lambda=2$ most significant bits of the addition result to perform approximate normalization.
In BF16 and BF16an-$k$-$\lambda$, the partial sums of each column of the matrix engine utilize twice the bit width of the input significands, as shown in Fig. \ref{f:mul-add-pe}. In all cases, activation functions are computed in FP32.

BF16an-1-1 and BF16an-1-2 exhibit 1\% degradation, on average, in accuracy and F1-score, while BF16an-2-2 performs worst and degrades accuracy by 7.2\% on average.
BF16an-1-1 and BF16an-1-2 retain higher accuracy than BF16an-2-2 because they can detect safely the most probable normalization shifts of $1$--$3$ bit positions, respectively. 
On the contrary, BF16an-2-2 leaves in most cases the result only partially normalized, which in turn causes the alignment of the subsequent PEs to introduce a significant amount of error since bits
are shifted out permanently, more often.

\begin{figure}
    \centering
    \includegraphics[width=0.9\columnwidth]{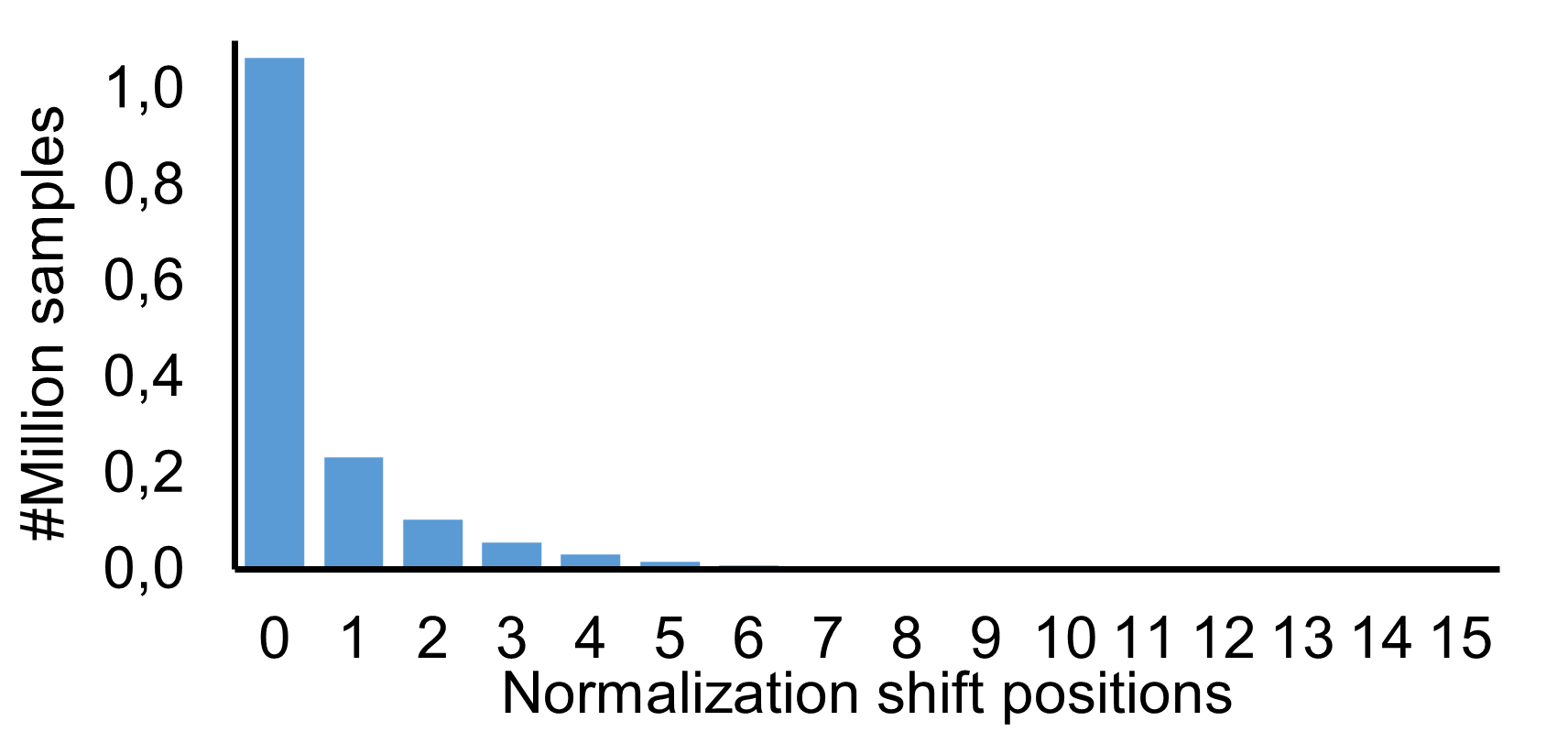}
    \caption{Histogram of the average number of normalization shifts for the matrix multiplications in three randomly selected attention layers of the BERT transformer.}
    \label{f:norm_dist}
\end{figure}

The higher performance offered by the case of $k=1$ and $\lambda=2$ can be easily explained by noticing that in BF16 that performs accurate normalization, the cases that needed more than three shifts are very few. This behavior is highlighted in 
Fig.~\ref{f:norm_dist} that depicts the average number of normalization shifts needed in the matrix multiplications of all attention layers of BERT.

\subsection{Area and Power evaluation}
To assess the area and power savings derived from approximate normalization for our most accurate model; BF16an-1-2, with respect to its BF16 counterpart that employs accurate normalization, we implemented the designs under comparison in SystemVerilog RTL and implemented them with Cadence’ digital implementation flow using a 28 nm standard-cell library for a target clock frequency of 1 GHz.

\begin{figure}
    \centering
    \includegraphics[width=0.98\columnwidth]{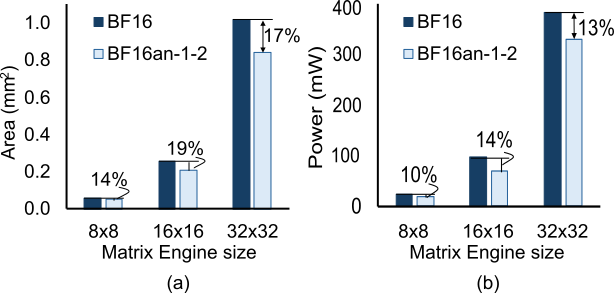}
    \caption{The total (a) area and (b) power savings of matrix engines that use approximate normalization for various sizes. Both diagrams highlight separately the contribution of approximate normalization.}
    \label{f:area_power_1GHz}
\end{figure}

Fig. \ref{f:area_power_1GHz} shows the area and power savings offered by using approximate normalization, for three different sizes of matrix engines. Power measurements were performed using the same data that were used for the inference tasks reported in Table \ref{t:accuracy}.
The results show that by employing approximate normalization, we can achieve area and power savings in the range of 14-19\% and 10-14\%, respectively.
\section{Conclusions}
Decades-old studies focusing on the behavior of floating-point addition with respect to normalization has shown that normalization rarely needs large shifts. In the majority of cases, to normalize the result of the addition a left shift by one or two positions, or an one-bit right shift in the case of addition overflow, is needed. In this work, we leverage this fundamental property of normalization to design low-cost floating-point multiply-add units that employ approximate normalization instead of an accurate one. The proposed approximate normalization reduced the hardware complexity of the floating-point multiply-add units of matrix engines, without ruining the accuracy of representative transformer-based applications.

\bibliographystyle{IEEEtran}
\bibliography{refs}

\end{document}